\documentclass[prd,aps,showpacs,preprintnumbers,amssymb]{revtex4}
\usepackage{axodraw}
\usepackage{color}
\usepackage{epsf}

\def\e3p{$\eta \rightarrow 3 \pi$}

\begin{document}
\title{%
\hfill{\normalsize\vbox{%
\hbox{}
 }}\\
{A  nonperturbative method for the scalar field theory }}

\author{Renata Jora
$^{\it \bf a}$~\footnote[2]{Email:
 rjora@theory.nipne.ro}}

\affiliation{$^{\bf \it a}$ National Institute of Physics and Nuclear Engineering PO Box MG-6, Bucharest-Magurele, Romania}

\date{\today}

\begin{abstract}
We compute an all order correction to the scalar mass in the $\Phi^4$ theory using a new method of  functional integration adjusted also to the large couplings regime.
\end{abstract}
\maketitle

\section{Introduction}

Currently very much is known about the perturbative behavior of many theories with or without gauge fields.  Beta functions for the $\Phi^4$ theory and QED is known up to the fifth order whereas for QCD is known up  to the fourth  order \cite{Vladimirov}-\cite{Baikov}. However there is limited knowledge  regarding the non-perturbative behavior of the same theories. Recently attempts \cite{Sannino} have been made for  determining the existence in some renormalization scheme of all order beta functions for gauge theories with various representations of fermions.
It is rather useful to search for alternative methods which may reveal either the higher orders of perturbation theories or even  the non-perturbative regime.

Here we shall consider the massive $\Phi^4$ theory as a laboratory for implementing a method  that can be further applied to more comprehensive models. There is an ongoing debate with regard to the behavior of the renormalized  coupling  $\lambda$  at small momenta referred to as "the triviality problem" \cite{Wilson}-\cite{Podolsky}.   With the hope that our approach might shed light even on this problem  we introduce a new variable in the path integral formalism  which allows for a more tractable functional integration and series expansion. Then we compute in this new method the corrections to the mass of the scalar in all order of perturbation theory.  This approach should be regarded as an alternate renormalization procedure. Since the corresponding mass anomalous dimension
$\gamma(m^2)=\frac{ d\ln m^2}{d\mu^2}$ has the first order (one loop) coefficient universal we verify that the first order correction is correct. However we expect that the next orders are different.

\section{ The set-up}

We shall illustrate our approach for a simple scalar theory, given by the Lagrangian:
\begin{eqnarray}
&&{\cal L}={\cal L}_0+{\cal L}_1
\nonumber\\
&&{\cal L}_0=\frac{1}{2}(\partial_{\mu}\Phi)(\partial^{\mu}\Phi)-\frac{1}{2}m_0^2\Phi^2
\nonumber\\
&&{\cal L}_1=-\frac{\lambda}{4!}\Phi^4.
\label{lagr5647}
\end{eqnarray}

We will work both in the Minkowski and Euclidian space upon convenience.

The generating functional in the euclidian  space has the expression:
\begin{eqnarray}
W[J]=
\int d\Phi\exp[-\int d^4x[\frac{1}{2}(\frac{\partial\Phi}{\partial\tau})^2+\frac{1}{2}(\Delta\Phi)^2+
\frac{1}{2}m_0^2\Phi^2+\frac{\lambda}{4!}\Phi^4+J\Phi]]
\label{gen5647}
\end{eqnarray}

and can be written as:
\begin{eqnarray}
W[J]=\exp[\int d^4x{\cal L}_1(\frac{\delta}{\delta J})]W_0[J]
\label{func65784}
\end{eqnarray}

where,
\begin{eqnarray}
W_0[J]=\int d\Phi\exp[\int d^4x({\cal L}_0+J\Phi)].
\label{funcz456}
\end{eqnarray}

From Eq.(\ref{func65784}) is clear how the perturbative approach can work. If $\lambda$ is a small parameter one can expand the exponential  in terms of $\lambda$ and solve succesive contributions accordingly.
 However we are interested in the regime where $\lambda$ is large  and one cannot use the above expansion.

We will illustrate our approach simply on a simple function. Assume we have the following one-dimensional integral which cannot be solved analytically:
\begin{eqnarray}
I=\int dx\exp[-af(x)],
\label{int5678}
\end{eqnarray}
where f is polynomial of x.
For $a$ small the expansion in $a$ makes sense. For $a\rightarrow\infty$ the  Taylor expansion uses:
\begin{eqnarray}
\lim_{a\rightarrow \infty}\frac{d^n \exp[-af(x)]}{da^n}=0
\label{wrontr786}
\end{eqnarray}
which does not lead to a correct answer.

We shall use however a simple trick.
We replace in the polynomial f some of the variables x with a new variable y (for example $x^4\rightarrow x^2y^2$). Then we write:
\begin{eqnarray}
I&&=\int dx dy\delta(x-y)exp[-af(x,y)]=\int dx dydz\exp[-i(x-y)z]\exp[-af(x,y)]=
\nonumber\\
&&\int dx dy dz \exp[-i(x-y)z-af(x,y)]
\label{rez67589}
\end{eqnarray}

This does not help too much in the present form. However if $f(x,y)=x^2y^2$ or any other function that contains $x^2$ we can form the perfect square:
\begin{eqnarray}
-ixz-ax^2y^2=-(\sqrt{a}xy+\frac{iz}{2\sqrt{a}y})^2-\frac{z^2}{4ay^2}.
\label{rez678}
\end{eqnarray}

Introduced in Eq. (\ref{rez67589}) this leads:
\begin{eqnarray}
I={\rm const}\int d\frac{1}{\sqrt{a}y} dz \exp[-\frac{z^2}{4ay^2}]\exp[iyz]
\label{rez3425}
\end{eqnarray}

Then expansion in $\frac{1}{a}$ makes sense and one can write:
\begin{eqnarray}
I={\rm const}\int dx dz\frac{1}{\sqrt{a}y}[1-\frac{z^2}{4ay^2}+...]\exp[iyz]
\label{rez4567}
\end{eqnarray}

This expansion may seem ill defined and highly divergent. For example if one integrates over z already encounters infinities. However in the functional method one is dealing with functions
instead of simple variables and  one encounters divergences also in the usual expansion in small parameters. Such that we will consider the above  approach as our starting point and
solve the problem of divergences as they appear.

We will start with the simple partition function for a $\Phi^4$ theory without a source:
\begin{eqnarray}
W[0]=\int d\Phi\exp[i\int d^4x [{\cal L}_0+{\cal L}_1]]
\label{def5678}
\end{eqnarray}

We consider the extended functional $\delta$ defined in the Minkowski space as (see the Appendix):
\begin{eqnarray}
\delta(\Phi)={\rm const} \int dK \exp[i\int d^4x_M K\Phi]
\label{rez5677}
\end{eqnarray}
which in the euclidian space becomes:
\begin{eqnarray}
\delta(\Phi)={\rm const}\int dK\exp[-\int d^4x K\Phi]
\label{rez453627}
\end{eqnarray}

We then rewrite Eq. (\ref{def5678})in Minkowski space as:
\begin{eqnarray}
&&W[0]=\int d\Phi d\Psi \delta(\Phi-\Psi)\exp[i\int d^4x [{\cal L}_0-\frac{\lambda}{8}\Phi^2\Psi^2]]=
\nonumber\\
&&{\rm const}\int d\Phi d\Psi d K\exp[i\int d^4x K(\Phi-\Psi)] \exp[i\int d^4x [{\cal L}_0-\frac{\lambda}{8}\Phi^2\Psi^2]]=
\nonumber\\
\nonumber\\
&&{\rm const}\int \frac{1}{\sqrt{\lambda}}d\Phi dK\exp[i\int d^4 x\frac{2}{\lambda}K^2]\exp[i\int d^4x K\Phi^2]\exp[i\int d^4x{\cal L}_0]
\label{eq4536}
\end{eqnarray}

In order to obtain this result we made the following  change of variable in the second line of Eq. (\ref{eq4536}): $K\rightarrow K\Phi$, $\Psi\rightarrow \frac{\Psi}{\Phi \sqrt{\lambda}}$.
Note that the $\lambda$ term gets rescaled by $3$ such that to take into account the various contribution of the Fourier modes.

We will estimate the first order of the integral in Eq. (\ref{eq4536}) given by:
\begin{eqnarray}
{\rm const}\int \frac{1}{\sqrt{\lambda}}d\Phi dK\frac{1}\exp[i\int d^4x K\Phi^2]\exp[i\int d^4x{\cal L}_0]
\label{int67589}
\end{eqnarray}

In order to solve the integral we write:
\begin{eqnarray}
&&\int d^4x [K\Phi^2+{\cal L}_0]=
\int d^4x [\frac{1}{V^2}\sum_{k_n+k_m+k_p=0}({\rm Re}K_m+i{\rm Im}K_m)({\rm Re}\Phi_n+i{\rm Im}\Phi_n)({\rm Re}\Phi_p+i{\rm Im}\Phi_p)-
\nonumber\\
&&-\frac{1}{2V}\sum_{k_n}(m_0^2-k_n^2)
[({\rm Re}\Phi_n)^2+({\rm Im}\Phi_n)^2]]
\label{exp8970}
\end{eqnarray}

We denote the bilinear form in the exponential in the Eq.(\ref{exp8970}) by:
\begin{eqnarray}
\Phi[\frac{K}{V^2}-\frac{1}{2V}[\frac{2K_0}{V}-(m_0^2-p_n^2)(\delta_{2n+1,2n+1}+\delta_{2n+2,2n+2})]]\Phi
\label{res6578}
\end{eqnarray}
where the counting starts from $n=0$ and we arranged for example the ${\rm Re}\Phi_n$ and ${\rm Im}\Phi_n$ components in the $2n+1$, respectively $2n+2$ columns of an infinitely dimensional vector.

Then the integral in Eq. (\ref{int67589}) can be solved easily as a  gaussian integral:
\begin{eqnarray}
&&{\rm const}\int \frac{1}{\sqrt{\lambda}}d\Phi dK\exp[i\int d^4x K\Phi^2]\exp[i\int d^4x{\cal L}_0]=
\nonumber\\
&&=\int d K \frac{1}{\det[\frac{K}{V^2}+\frac{1}{2V}[\frac{2K_0}{V}-(m_0^2-p_n^2)(\delta_{2n+1,2n+1}+\delta_{2n+2,2n+2})]]^{1/2}}.
\label{res6578349}
\end{eqnarray}

Note that one can write also a result for the full partition function in Eq.  (\ref{eq4536}):
\begin{eqnarray}
&&{\rm const}\int \frac{1}{\sqrt{\lambda}}d\Phi dK\exp[i\int d^4 x\frac{2}{\lambda}K^2]\exp[i\int d^4x K\Phi^2]\exp[i\int d^4x{\cal L}_0]=
\nonumber\\
&&=\int d K \exp[i\int d^4 x\frac{2}{\lambda}K^2]\frac{1}{\det[\frac{K}{V^2}+\frac{1}{2V}[\frac{2K_0}{V}-(m_0^2-p_n^2)(\delta_{2n+1,2n+1}+\delta_{2n+2,2n+2})]]^{1/2}}.
\label{fullpart5467}
\end{eqnarray}

The next step is to determine through this procedure the propagator.
\section{The propagator}

The propagator is given by:
\begin{eqnarray}
\langle\Omega|T\Phi(x_1)\Phi(x_2)|\Omega\rangle=
\lim_{T\rightarrow \infty}\frac{ \int d\Phi\Phi(x_1)\Phi(x_2)\exp[i\int^T_{-T}d^4 x{\cal L}]}{\int d\Phi\exp[i\int^T_{-T}d^4 x{\cal L}]}.
\label{formula67}
\end{eqnarray}

For our partition function the Eq. (\ref{formula67}) is rewritten as:
\begin{eqnarray}
&&\langle\Omega|T\Phi(x_1)\Phi(x_2)|\Omega\rangle=
\frac{\int \frac{1}{\sqrt{\lambda}}d\Phi d K \Phi(x_1)\Phi(x_2)\exp[i\int d^4x\frac{2}{\lambda}K^2]\exp[i\int d^4x K\Phi^2]\exp[i\int d^4x{\cal L}_0]}
{\int \frac{1}{\sqrt{\lambda}}d\Phi d K \exp[i\int d^4x\frac{2}{\lambda}K^2]\exp[i\int d^4x K\Phi^2]\exp[i\int d^4x{\cal L}_0]}=
\nonumber\\
&&\frac{1}{V^2}\sum_m \exp[-i p_m(x_1-x_2)]iV\frac{\frac{\delta}{\delta(m^2-p_m^2)}\int d\Phi d K \exp[i\int d^4x\frac{2}{\lambda}K^2]\exp[i \int d^4xK\Phi^2]\exp[i d^4x {\cal L}_0]}
{\int d\Phi d K \exp[i\int d^4x\frac{2}{\lambda}K^2]\exp[i \int d^4xK\Phi^2]\exp[i d^4x {\cal L}_0]}
\label{firstrez45367}
\end{eqnarray}

Note that the first line in Eq. (\ref{firstrez45367}) is the standard definition of the two point function. The second line in Eq. (\ref{firstrez45367}) needs some clarification. From the first line in the equation it can be seen that the scalar two point function may receive contributions either from the kinetic term or from the terms that contain K. We need to show that also the second line is justified.  In order to see that  one should consider the simple functional integral in Eq. (\ref{firstrez45367}) and treat it independently without any reference to the Feynman diagrams. Then the first line of Eq. (\ref{firstrez45367}) leads also to:
\begin{eqnarray}
&&\langle\Omega|T\Phi(x_1)\Phi(x_2)|\Omega\rangle=
\frac{\int \frac{1}{\sqrt{\lambda}}d\Phi d K \Phi(x_1)\Phi(x_2)\exp[i\int d^4x\frac{2}{\lambda}K^2]\exp[i\int d^4x K\Phi^2]\exp[i\int d^4x{\cal L}_0]}
{\int \frac{1}{\sqrt{\lambda}}d\Phi d K \exp[i\int d^4x\frac{2}{\lambda}K^2]\exp[i\int d^4x K\Phi^2]\exp[i\int d^4x{\cal L}_0]}=
\nonumber\\
&&\frac{1}{V^2}\sum_m\sum_n\exp[-ip_m x_1]\exp[-ip_n x_2]iV^2\frac{\int d\Phi d K \exp[i\int d^4x\frac{2}{\lambda}K^2]\frac{\delta}{\delta K(-p_m-p_n)}\exp[i \int d^4xK\Phi^2]\exp[i d^4x {\cal L}_0]}
{\int d\Phi d K \exp[i\int d^4x\frac{2}{\lambda}K^2]\exp[i \int d^4xK\Phi^2]\exp[i d^4x {\cal L}_0]}=
\nonumber\\
&&\frac{1}{V^2}\sum_m\sum_n\exp[-ip_m x_1]\exp[-ip_n x_2](-i)V^2\frac{\int d\Phi d K \frac{\delta}{\delta K(-p_m-p_n)}\exp[i\int d^4x\frac{2}{\lambda}K^2] \exp[i \int d^4xK\Phi^2]\exp[i d^4x {\cal L}_0]}
{\int d\Phi d K \exp[i\int d^4x\frac{2}{\lambda}K^2]\exp[i \int d^4xK\Phi^2]\exp[i d^4x {\cal L}_0]}=
\nonumber\\
&&\frac{1}{V^2}\sum_m\sum_n\exp[-ip_m x_1]\exp[-ip_n x_2](-\frac{2}{\lambda})V\frac{\int d\Phi d K K(p_n+p_m)\exp[i\int d^4x\frac{2}{\lambda}K^2] \exp[i \int d^4xK\Phi^2]\exp[i d^4x {\cal L}_0]}
{\int d\Phi d K \exp[i\int d^4x\frac{2}{\lambda}K^2]\exp[i \int d^4xK\Phi^2]\exp[i d^4x {\cal L}_0]}
\label{rez54678}
\end{eqnarray}
We shall attempt to estimate the integral over the modes K(p) in the Eq. (\ref{rez54678}).  For that we expand the exponential of the trilinear term. In first order we get the term,
\begin{eqnarray}
K(p_m+p_n)K(-q-r)\Phi(q)\Phi(r)
\label{first567}
\end{eqnarray}
which is evident that brings contribution only for $q=-r$, $p_m=-p_n$ so the only K mode that contributes is the zero mode. In third order order (second order is zero) we obtain terms of the type:
\begin{eqnarray}
K(p_m+p_n)K(-q_1-r_1)K(-q_2-r_2)K(-q_3-r_3)\Phi(q_1)\Phi(r_1)\Phi(q_2)\Phi(r_2)\Phi(q_3)\Phi(r_3)
\label{arr4567}
\end{eqnarray}
If any of the $q_i=-r_i$ we are back to the previous case where only $K_0$ mode contribute. Assume without loss of generality that $q_1=-q_2$, $r_1=-r_3$, $q_3=-r_2$.  This settle the integral over $\Phi$ whereas for $K$ we obtain:
\begin{eqnarray}
K(p_m+p_n)K(-q_1-r_1)K(q_1-r_2)K(r_1+r_2)
\label{termofinterset}
\end{eqnarray}
There are three possibilities for this integral: 1) $p_m+p_n=q_1+r_1$, $q_1-r_2=-r_1-r_2$, 2) $p_m+p_n=-q_1+r_2$, $q_1+r_1=r_1+r_2$, 3) $p_m+p_n=-r_1-r_2$, $q_1+r_1=q_1-r_2$. All of these possibilities lead to $p_n=-p_m$. This arguments  continues for higher orders in the expansion such that quite justified we can express the propagator from the beginning as the derivative with respect to $m^2-p_m^2$.


Since the quantity $m^2-k_m^2$ appears only in the determinant in Eq. (\ref{fullpart5467}) we can compute:
\begin{eqnarray}
&&\frac{\delta}{\delta (m_0^2- p_m^2)}  [\det[\frac{K}{V^2}+\frac{1}{2V}[\frac{2K_0}{V}-(m_0^2-p_n^2)(\delta_{2n+1,2n+1}+\delta_{2n+2,2n+2})]]]^{-1/2}=
\nonumber\\
&&-\frac{1}{2}[\det[\frac{K}{V^2}-\frac{1}{2V}[\frac{2K_0}{V}-(m_0^2-p_n^2)(\delta_{2n+1,2n+1}+\delta_{2n+2,2n+2})]]]^{-1/2}\times
\nonumber\\
&&Tr[\frac{1}{\frac{K}{V^2}+
\frac{1}{2V}[2\frac{K_0}{V}-(m_0^2-p_n^2)(\delta_{2n+1,2n+1}+\delta_{2n+2,2n+2})]}(-1)(\frac{1}{2V}(\delta_{2m+1,2m+1}+\delta_{2m+2,2m+2}))]=
\nonumber\\
&&{\rm const} \frac{1}{2}[\det[\frac{K}{V^2}-\frac{1}{2V}[2\frac{K_0}{V}-(m_0^2-p_n^2)(\delta_{2n+1,2n+1}+\delta_{2n+2,2n+2})]]]^{-1/2}\frac{2}{\frac{2}{V}K_0-(m_0^2-p_m^2)}=
\nonumber\\
&&{\rm const}[\det[\frac{K}{V^2}-\frac{1}{2V}[2K_0-(m_0^2-p_n^2)(\delta_{2n+1,2n+1}+\delta_{2n+2,2n+2})]]]^{-1/2}\frac{1}{\frac{2K_0}{V}-(m_0^2-p_m^2)}.
\label{dif657}
\end{eqnarray}

In Eq. (\ref{dif657}) the first three lines are the simple result of differentiating a determinant. The first factor in the third line of Eq. (\ref{dif657}) contains the Fourier modes of  the field K with momenta different than zero ($p_{\mu}\neq 0$) denoted simply by K and those with momenta $p_{\mu}=0$ denoted by $K_0$. However the modes with $p_{\mu}\neq 0$ are irrelevant for the reason we shall outline below. First let us consider $K(x)$ as a square integrable function in the Hilbert space which satisfies:
\begin{eqnarray}
\int d^4x K^2(x)=\frac{1}{V}\sum_p K(p)^2<M,
\label{res5467}
\end{eqnarray}
where M is a quantity large but finite. This means that $\frac{K(p)}{V}<\frac{\sqrt{M}}{\sqrt{V}}$. In contrast $\frac{K_0}{V}$ is finite as is given by:
\begin{eqnarray}
\frac{K_0}{V}=\int d^4 x K(x)
\label{res54678}
\end{eqnarray}
We could have dropped from the beginning the factor $\frac{1}{V}$ from its expression but it helps with dimensional analysis. Thus although we shall keep $K(p\neq 0)$ in the expression at some point the limit $V\rightarrow \infty$ will be taken such that all these terms in the determinant will cancel and the integral of the exponential of the $K(p_n)$ terms in the numerator will get canceled by that in the denominator.  In conclusion the zero mode is used as a substitute for all the the other modes and sums up all their contribution.

The mode $K_0$ acts like an additional contribution to the scalar mass and needs to be maintained and integrated over. In consequence in all calculations that follows one should consider only the modes $K_0$ facts which simplifies the calculations considerably.

Then Eq. (\ref{firstrez45367}) becomes:
\begin{eqnarray}
&&\langle\Omega|T\Phi(x_1)\Phi(x_2)|\Omega\rangle=
\frac{1}{V^2}\sum_m \exp[-i p_m(x_1-x_2)]iV\times
\nonumber\\
&&\times\frac{\int d K \frac{1}{\frac{2}{V}K_0-(m_0^2-p_m^2)}\exp[i\int d^4x \frac{2}{\lambda}K^2]\frac{1}{\det[\frac{K}{V^2}+\frac{1}{2V}[2\frac{K_0}{V}-(m_0^2-p_n^2)(\delta_{2n+1,2n+1}+\delta_{2n+2,2n+2})]]^{1/2}}}
{\int d K \exp[i\int d^4x \frac{2}{\lambda}K^2]\frac{1}{\det[\frac{K}{V^2}+\frac{1}{2V}[\frac{2K_0}{V}-(m_0^2-p_n^2)(\delta_{2n+1,2n+1}+\delta_{2n+2,2n+2})]]^{1/2}}}
\label{rez45367}
\end{eqnarray}

We denote:
\begin{eqnarray}
&&\frac{1}{\lambda}\frac{2}{V}=ba_0
\nonumber\\
&&m_0^2-p_m^2=c^2
\nonumber\\
&&\frac{2}{V}=a_0
\nonumber\\
&&\det[\frac{K}{V^2}+\frac{1}{2V}[\frac{2K_0}{V}-(m_0^2-p_n^2)(\delta_{2n+1,2n+1}+\delta_{2n+2,2n+2})]]=\det[a_0K_0+B]
\label{not67589}
\end{eqnarray}

We need to evaluate:
\begin{eqnarray}
&&\frac{\int d K d K_0\exp[2iba_0 K_0^2 +\int d^4x 2ib K^2]\frac{1}{(a_0K_0-c^2)[\det[a_0K_0+B]]^{1/2}}}{\int d K d K_0\exp[ib K_0^2+\int d^4x ibK^2]\frac{1}{[\det[a_0K_0+B]]^{1/2}}}=
\nonumber\\
&&-\frac{1}{c^2}\frac{\int d K d K_0\exp[2i ba_0 K_0^2+\int d^4x 2ibK^2][1+\frac{a_0K_0}{c^2}+\frac{a_0^2K_0^2}{c^4}+...]\frac{1}{[\det[a_0K_0+B]]^{1/2}}}{\int d K d K_0\exp[i b K_0^2+\int d^4x ibK^2]\frac{1}{[\det[a_0K_0+B]]^{1/2}}}
\label{rez56478}
\end{eqnarray}

We extracted a factor of $\frac{1}{V}$ from the determinant and dropped the corresponding constant factor everywhere.
In order to determine the ratio in Eq. (\ref{rez56478}) we evaluate each term in the expansion in the denominator:
\begin{eqnarray}
&&I_{n}=\int d K_0 d K \frac{(a_0K_0)^{n}}{c^{2n}}\exp[2i b a_0K_0^2]\exp[\int d^4 x 2 i bK^2](\det[a_0K_0+B])^{-1/2}=
\nonumber\\
&&\int d K_0 d K\frac{1}{a_0}\frac{d[\frac{(a_0K_0)^{n+1}}{c^{2n}(n+1)}]}{d K_0}\exp[2i ba_0 K_0^2]\exp[\int d^4x 2i bK^2](\det[a K_0+B])^{-1/2}=
\nonumber\\
&&-\int d K_0 d K \frac{1}{a_0}\frac{(a_0K_0)^{n+1}}{c^{2n}(n+1)}(4iba_0K_0)\exp[2i b a_0K_0^2]\exp[\int d^4x 2 i bK^2](\det[a K_0+B])^{-1/2}-
\nonumber\\
&&\int d K_0 d K \frac{(a_0K_0)^{n+1}}{a_0c^{2n}(n+1)}\exp[2i b a_0K_0^2]\sum_k[\frac{-a_0}{a_0K_0-c_k^2}]\exp[\int d^4x 2i bK^2](\det[a_0K_0+B])^{-1/2}=
\nonumber\\
&&-\frac{4ibc^4}{a_0(n+1)}I_{n+2}-\int d K d K_0\frac{(a_0K_0)^{n+1}}{c^{2n}(n+1)}\sum_k\frac{1}{c_k^2}[1+\frac{a_0K_0}{c_k^2}+\frac{(a_0K_0)^2}{c_k^4}+...]\times
\nonumber\\
&&\exp[2i ba_0 K_0^2]\exp[\int d^4x 2i bK^2](\det[a_0K_0+B])^{-1/2}.
\label{rez54678}
\end{eqnarray}
Here we used the formula of differentiation of a determinant.

From Eqs. (\ref{rez54678}) and (\ref{rez5678}) we obtain the following recurrence formula:
\begin{eqnarray}
(n+1)I_n+I_{n+2}c^4[\frac{4ib}{a_0}+\sum_k \frac{1}{c_k^4}]+I_{n+1}c^2\sum_k\frac{1}{c_k^2}+...+I_{n+r}c^{2r}\sum_k\frac{1}{c_k^{2r}}+...=0
\label{rec345}
\end{eqnarray}

First we multiply the whole  Eq. (\ref{rec345}) by $\frac{1}{V}$ and then introduce $I_{n}c^{2n}=J_n$ to get the new recurrence formula:
\begin{eqnarray}
\frac{1}{V}(n+1)J_n+J_{n+1}\frac{1}{V}\sum_k\frac{1}{c_k^2}+J_{n+2}[2ib+\frac{1}{V}\sum_k\frac{1}{c_k^4}]+...=0
\label{new789}
\end{eqnarray}

Finally since we denoted the partition function by $I_0$ from Eqs. (\ref{rez45367}) and (\ref{new789}) one can derive:
\begin{eqnarray}
&&{\rm Propagator}=-\frac{1}{c^2}\sum_n I_n/I_0=
\nonumber\\
&&=-\frac{1}{c^2}\sum_n\frac{1}{c^{2n}}J_n/I_0,
\label{finalt657}
\end{eqnarray}
where $J_0=I_0$ is the full partition function.

Before going further we need to determine the coefficients in Eq. (\ref{new789}). For that we first state,
\begin{eqnarray}
\frac{1}{V}\sum_k\frac{1}{c_k^{2r}}=\frac{1}{V}\sum_k\frac{1}{(m_0^2-p_k^2)^r}=
(-1)^r\int d^4 p\frac{1}{(p^2-m_0^2)^r}=q_r.
\label{rez5647}
\end{eqnarray}

Note that only the integral with $k=1,2$ are divergent whereas the other ones are finite. We shall use a simple cut-off the regularize them upon the case. Then we get:
\begin{eqnarray}
&&q_1=i\frac{1}{16\pi^2}[\Lambda^2-m_0^2\ln[\frac{\Lambda^2}{m_0^2}]]
\nonumber\\
&&q_2=i\frac{1}{16\pi^2}[-1+\ln[\frac{\Lambda^2}{m_0^2}]]
\nonumber\\
&&q_{n,n>2}=i\frac{1}{16\pi^2}\frac{(m_0^2)^{2-n}}{(n-1)(n-2)}.
\label{finrez45467}
\end{eqnarray}

\section{Discussion and conclusions}

The terms $J_n$ in the two point function in Eq. (\ref{finalt657}) correspond to various loop corrections and one can cut the series to obtain results in various orders of perturbation theory. However we shall not attempt to do this here. We will rather aim to obtain if possible an all order result for the correction to the mass of the scalar. We do this with the hope that the approach initated here can be extended easily to theories with spontaneous symmetry breaking and even to the standard model. It is clear that an approach that could estimate the correction to the Higgs boson mass could prove of great interest.
One can write quite generally an exact expression for the propagator of a scalar:
\begin{eqnarray}
\frac{i}{p^2-m^2-M^2(p^2)}
\label{prop6578}
\end{eqnarray}
where $m$ is the physical mass and $M^2(p^2)$ is the one particle irreducible self energy (For simplicity we rename $p_m^2=p^2$ for the rest of the paper). In our approach the propagator is given by:
\begin{eqnarray}
\frac{i}{p^2-m_0^2}\sum_n(-1)^n\frac{J_n}{I_0}\frac{1}{(p^2-m_0^2)^n}
\label{rez5678}
\end{eqnarray}

Now if we identify Eq. (\ref{prop6578}) with Eq. (\ref{rez5678}) and expand the first equation in series in $\frac{1}{(p^2-m_0^2)^n}$ we obtain :
\begin{eqnarray}
&&\frac{i}{p^2-m_0^2}\sum_n(-1)^n\frac{J_n}{I_0}\frac{1}{(p^2-m_0^2)^n}=\frac{i}{p^2-m^2-M^2(p^2)}
\nonumber\\
&&(-1)^n\frac{J_n}{I_0}=[m^2-m_0^2-M^2(p^2)]^n+Y_n(p^2)
\nonumber\\
&&\frac{J_n}{I_0}=[m_0^2-m^2-M^2(p^2)]^n+(-1)^nY_n(p^2),
\label{ident6578}
\end{eqnarray}
where $Y_n(p^2)$ is an arbitrary series with the property:
\begin{eqnarray}
\sum_n Y_n(p^2)\frac{1}{(p^2-m_0^2)^{n+1}}=0
\label{rez56478}
\end{eqnarray}

Now we shall consider the following  renormalization conditions which states:
\begin{eqnarray}
&&M^2(p^2)_{p^2=m^2}=0
\nonumber\\
&&\frac{d M^2(p^2)}{dp^2}|_{p^2=m^2}=0
\label{renorm8970}
\end{eqnarray}

We apply the first condition to Eq. (\ref{ident6578}) to determine that:
\begin{eqnarray}
\frac{(-1)^n J_n}{I_0}|_{p^2=m^2}=(m^2-m_0^2)^n+Y_n(m^2)=(m^2-m_0^2)^n(1+\frac{Y_n}{(m^2-m_0^2)^n})
\label{condre456}
\end{eqnarray}

We will show that the term $Y_n(m^2)$ in Eq. (\ref{condre456}) should be set to zero. For that we first note that from Eq. (\ref{rez56478}) one can deduce that there is at least one n for which $\frac{Y_n(m^2)}{(m^2-m_0^2)^n}< 0$. Then there is a solution $m$ for which $(m^2-m_0^2)^n=-\frac{1}{Y_n(m^2)}=\alpha(n)^n$ with $\alpha(n)$ real. This solution is a zero of the corresponding $J_n$. But $J_n(m^2)$ has the expression:
\begin{eqnarray}
\int d K_0 K_0^n\frac{1}{a_0K_0-\alpha(n)}\times{\rm other \,factors},
\label{somer456}
\end{eqnarray}
so has a pole at $a_0K_0=\alpha(n)$ instead of a zero. We obtain a contradiction which means that  there is no $n$ such that $\frac{Y_n(m^2)}{(m^2-m_0^2)^n}< 0$  so the series in Eq. (\ref{rez56478}) has all the terms $Y_n(m^2)=0$.




Then we simply take:
\begin{eqnarray}
\frac{J_n}{I_0}|_{p^2=m^2}=[m_0^2-m^2]^n.
\label{final675}
\end{eqnarray}

We  denote,
\begin{eqnarray}
X=[m_0^2-m^2],
\label{not7689}
\end{eqnarray}

and sum in the recurrence formula in Eq. (\ref{new789}) all terms with the indices $n+k$, $k\geq 3$ for $p^2=m^2$.
\begin{eqnarray}
&&\sum_{k\geq 3}\frac{J_{n+k}}{I_0}q_k=X^n\sum_k \frac{i}{16\pi^2}m_0^4(\frac{X}{m_0^2})^n\frac{1}{(n-1)(n-2)}=
\nonumber\\
&&\frac{i}{16\pi^2}X^{n+1}[X+(m_0^2-X)\ln[\frac{m_0^2-X}{m_0^2}]]
\label{sum567489}
\end{eqnarray}

Then the recurrence formula becomes:
\begin{eqnarray}
&&(n+1)a_0X^n+q_1X^{n+1}+(\frac{2i}{\lambda}+q_2)X^{n+2}+\frac{i}{16\pi^2}X^{n+1}[X+(m_0^2-X)\ln[\frac{m_0^2-X}{m_0^2}]]=0
\nonumber\\
&&(n+1)a_0\frac{1}{X}+q_1+(\frac{2i}{\lambda}+q_2)X+\frac{i}{16\pi^2}[X+(m_0^2-X)\ln[\frac{m_0^2-X}{m_0^2}]]=0
\nonumber\\
&&q_1+(\frac{2i}{\lambda}+q_2)X+\frac{i}{16\pi^2}[X+(m_0^2-X)\ln[\frac{m_0^2-X}{m_0^2}]]=0.
\label{rez45367}
\end{eqnarray}
Here in the last line we took the limit $a_0=\frac{1}{V}\rightarrow0$.

Note that although we used the conditions in Eq. (\ref{renorm8970}) we should not consider our approach equivalent with any of the standard renormalization procedures.

Then Eq. (\ref{rez45367}) will become:
\begin{eqnarray}
q_1+(m_0^2-m^2)[\frac{2i}{\lambda}+q_2]+\frac{i}{16\pi^2}[(m_0^2-m^2)+m^2\ln[\frac{m^2}{m_0^2}]]=0
\label{rel7869}
\end{eqnarray}

The equation above determines the physical mass in terms of the bare mass and of the cut-off scale.
Instead we observe that for a large cut-off scale one can divide the Eq. (\ref{rel7869}) by $q_1$ and retain the first and second term. Then,
\begin{eqnarray}
m^2\approx m_0^2+\frac{q_1}{\frac{2i}{\lambda}+q_2}\approx
m_0^2+\frac{\Lambda^2-m_0^2\ln[\frac{\Lambda^2}{m_0^2}]}{1+\frac{\lambda}{32\pi^2}[-1+\ln[\frac{\Lambda^2}{m_0^2}]]}\frac{\lambda}{32\pi^2}.
\label{finaform90}
\end{eqnarray}

Note that this result leads to the same first order coefficient of the mass anomalous dimension as in  the standard renormalization procedures.

\section*{Acknowledgments} \vskip -.5cm
The work of R. J. was supported by a grant of the Ministry of National Education, CNCS-UEFISCDI, project number PN-II-ID-PCE-2012-4-0078.

\appendix
\section{}

In the following we will show that the relation in Eq (\ref{rez453627}) make sense perfect sense in the functional approach.
We start with:
\begin{eqnarray}
&&\int d K \exp[-\int d^4x K\Phi]=
\prod_{k_n^0>0}\int d{\rm Re}K_n d{\rm Im}K_n\exp[-\frac{1}{V}({\rm Re}\Phi_n+i{\rm Im}\Phi_n)({\rm Re}K_n+i{\rm Im}K_n)]=
\nonumber\\
&&=\prod_{k_n^0>0}\int  d{\rm Re}K_n d{\rm Im}K_n\exp[-\frac{1}{V}{\rm Re}\Phi_n{\rm Re}K_n-\frac{i}{V}{\rm Im}\Phi_n{\rm Re}K_n]
\exp[\frac{1}{V}{\rm Im}\Phi_n{\rm Im}K_n-\frac{i}{V}{\rm Re}\Phi_n{\rm Im}K_n].
\label{exp8790}
\end{eqnarray}

Next let us consider a regular integral of the type:
\begin{eqnarray}
&&\int dx \exp[-i p x-ap]dx=
\int dx [1-(ap)+\frac{1}{2}(ap)^2+...]\exp[-ipx]=
\nonumber\\
&&\int d x[1-(-i)a\frac{\delta}{\delta x}+a^2\frac{1}{2}(-i)^2\frac{\delta^2}{\delta x^2}+...]\exp[-ipx]=
\nonumber\\
&&[1-(-i)a\frac{\delta}{\delta x}+a^2\frac{1}{2}(-i)^2\frac{\delta^2}{\delta x^2}+...]\delta(x)=g(x)
\label{rez5467788}
\end{eqnarray}

Let us apply this result  to Eq. (\ref{exp8790}) with  the variable a replaced depending on the case by ${\rm Re}\Phi_n$ or ${\rm Im}\Phi_n$ :
\begin{eqnarray}
&&\int d K d\Phi f(\Phi)\exp[-\int d^4x K\Phi]=
{\rm const}\int\prod_{k_n^0>0}d{\rm Re }\Phi_n d{\rm Im}\Phi_n f({\rm Re}\Phi_k,{\rm Im}\Phi_k) \times
\nonumber\\
&&(1-{\rm Re}\Phi_n(-i)\frac{\delta}{\delta {\rm Im}}+...)\delta({\rm Im}\Phi_n)\times(1-{\rm Im}\Phi_n
(-i)\frac{\delta}{\delta{\rm Re}\Phi_n}+...)\delta({\rm Re}\Phi_n)
={\rm const}f(0,0)
\label{exp87906}
\end{eqnarray}

We will prove that by considering  a few terms in the above expansion. The zeroth order term contains two delta functions and clearly leads to $f(0,0)$.
 Another possible term is:
\begin{eqnarray}
&&\int \prod_{k_n^0>0}d{\rm Re }\Phi_n d{\rm Im}\Phi_n f({\rm Re}\Phi_k,{\rm Im}\Phi_k){\rm Re}\Phi_n\frac{\delta}{\delta{\rm Im}\Phi_n}\delta({\rm Im}\Phi_n)\delta({\rm Re}\Phi_n)=0
\label{term456}
\end{eqnarray}
by virtue of the $\delta({\rm Re}\Phi_n)$ function.
 Another possible term is,
 \begin{eqnarray}
 &&\int\prod_{k_n^0>0}d{\rm Re }\Phi_n d{\rm Im}\Phi_n f({\rm Re}\Phi_k,{\rm Im}\Phi_k){\rm Re}\Phi_n\frac{\delta}{\delta{\rm Im}\Phi_n}\delta({\rm Im}\Phi_n)
 {\rm Im}\Phi_n\frac{\delta}{\delta {\rm Re}\Phi_n}\delta({\rm Re}\Phi_n)=
 \nonumber\\
 &&-\prod_{k_n^0>0}{\rm Re }\Phi_n {\rm Im}\Phi_n \frac{\delta}{\delta {\rm Re}\Phi_n}[f({\rm Re}\Phi_k,{\rm Im}\Phi_k){\rm Re}\Phi_n\frac{\delta}{\delta{\rm Im}\Phi_n}\delta({\rm Im}\Phi_n)
 {\rm Im}\Phi_n\frac{\delta}{\delta {\rm Re}\Phi_n}]\delta({\rm Re}\Phi_n)=
 \nonumber\\
 &&-\prod_{k_n^0>0}{\rm Re }\Phi_n {\rm Im}\Phi_n{\rm Im}\Phi_n\frac{\delta}{\delta{\rm Im}\Phi_n}\delta({\rm Im}\Phi_n)f(0,{\rm Im}\Phi_n)=f(0,0)
 \label{rez5647899}
 \end{eqnarray}

It can be shown that all other terms are either zero or proportional to $f(0,0)$ which concludes our proof that the integral in Eq. (\ref{exp87906}) gives a well defined delta function.

\section{}

We shall present here an approximate estimate of the propagator for an arbitrary regularization scheme for the limit of large coupling $\lambda$.
We start from the recurrence relation in Eq. (\ref{new789}) which we rewrite here for completeness:
\begin{eqnarray}
\frac{1}{V}(n+1)J_n+J_{n+1}\frac{1}{V}\sum_k\frac{1}{c_k^2}+J_{n+2}[2ib+\frac{1}{V}\sum_k\frac{1}{c_k^4}]+...=0
\label{new78923}
\end{eqnarray}

We denote:
\begin{eqnarray}
\frac{1}{V}\sum_k\frac{1}{(c_k^2)^n}=s_n,
\label{rez45678}
\end{eqnarray}
where $s_n$ may be considered in any regularization scheme. First we will make a change of variable $K_0=\frac{K_0'}{\sqrt{\lambda}}$ and rewrite $J_n=\frac{S_n}{\lambda^{n/2}}$ where $S_n$ represent the same quantity as $J_n$ this time in the variable $K_0'$. Note that with this change of notation for $\lambda$ very large the  factor $\exp[2iba_0\frac{K_0^{\prime2}}{\lambda}]$ in Eq. (\ref{rez54678}) becomes negligible.

The recurrence formula becomes in terms of $S_n$:
\begin{eqnarray}
\frac{1}{V}(n+1)S_n/\lambda^{n/2}+S_{n+1}/\lambda^{(n+1)/2}s_1+S_{n+2}/\lambda^{(n+2)/2}[2ib+s_2]+S_{n+3}/\lambda^{(n+3)/2}s_3...=0
\label{newrec43567}
\end{eqnarray}

For large V and $\lambda$ one obtains in first order:
\begin{eqnarray}
S_{n+2}=-S_{n+1}\frac{\sqrt{\lambda}s_1}{2ib+s_2}
\label{rec3434}
\end{eqnarray}
to determine $S_n=(-1)^{n-1}S_1[\frac{\sqrt{\lambda}s_1}{2ib+s_2}]^{n-1}=(-1)^{n-1}x^{n-1}\lambda^{(n-1)/2}S_1$.

In order to determine $S_1$ we consider the zeroth order recurrence relation:
\begin{eqnarray}
\frac{1}{V}S_0+\sum_{n=3}S_1(-1)^{n-1}x^{n-1}\lambda^{(n-1)/2}/\lambda^{n/2}s_{n}=0
\label{newform5678}
\end{eqnarray}
This yields:
\begin{eqnarray}
S_1=-\frac{1}{V}\sqrt{\lambda}S_0/[\sum_{n=3}(-1)^{n-1}x^{n-1}s_{n}]
\label{form6789}
\end{eqnarray}

According to Eq. (\ref{finalt657}) the propagator is given by:
\begin{eqnarray}
&&{\rm Propagator}=-\frac{1}{c^2}\sum_n\frac{1}{c^{2n}}(S_n\lambda^{-n/2})/S_0=
\nonumber\\
&&=-\frac{1}{c^2}[S_0+\sum_{n=1}\frac{1}{c^{2n}}(-1)^{n-1}x^{n-1}\lambda^{(n-1)/2}\lambda^{-n/2}S_1]/S_0=-\frac{1}{c^2}[S_0+\frac{1}{\sqrt{\lambda}(c^2+x)}S_1]/S_0=
\nonumber\\
&&-\frac{1}{c^2}[1-\frac{1}{c^2+x}\frac{1}{V[\sum_{n=3}(-1)^{n-1}x^{n-1}s_{n}]}],
\label{finalt65789}
\end{eqnarray}
where $x=\frac{s_1}{2ib+s_2}$ (See the notation in Eq. (\ref{not67589})). A straightforward computation for the quantities in Eq. (\ref{finalt65789}) in the limit of large $\lambda$ leads to:
\begin{eqnarray}
{\rm Propagator}\approx-\frac{1}{c^2}[1-\Lambda_1^2\frac{1}{c^2+\Lambda^2}],
\label{rez44567}
\end{eqnarray}
where $\Lambda_1^2={\rm function\,of} (\Lambda^2)<\Lambda^2$. Note that in the notation in the paper $c^2=m_0^2-p^2$.

 This is a particular case of " triviality" known to be a feature of the $\Phi^4$ theories in the limit where $\lambda\rightarrow\infty$.  To show this we rewrite the Eq. (\ref{rez44567}) as:
\begin{eqnarray}
&&{\rm Propagator}=\frac{1}{p^2-m_0^2}[1-\Lambda_1^2\frac{1}{m_0^2+\Lambda^2-p^2}]=
\nonumber\\
&&=\frac{\Lambda^2-\Lambda_1^2}{\Lambda^2}\frac{1}{p^2-m_0^2}+\frac{\Lambda_1^2}{\Lambda^2}\frac{1}{p^2-m_0^2-\Lambda^2}
\label{finlares6478}
\end{eqnarray}

According to \cite{Frasca} a theory is trivial in the strong coupling regime if the propagator can be written as:
\begin{eqnarray}
{\rm Propagator}=\sum_n\frac{Z_n}{p^2-m_n^2}
\label{def5467}
\end{eqnarray}
where $Z_n$ (all of them can be zero except one) are the weights and $m_n$ are the spectrum in the large coupling limit.
As it can be observed easily our result in Eq. (\ref{finlares6478}) is a particular case of triviality with the masses $m_1^2=m_0^2$ and $m_2^2=m_0^2+\Lambda^2$.

\end{document}